\documentstyle[12pt]{article}
\begin{document}
\begin{center}
{\bf Generalized Quantum Control-Not Gate in Two-Spin Ising System}\\ \ \\
Gennady P. Berman, Gary D. Doolen\\ 
Theoretical Division, 
T-13, MS B213, Los Alamos National Laboratory,\\ Los Alamos NM 87545 \\ \ \\
Gustavo V. L\'opez,\\
 Department of Physics, University of Guadalajara, Mexico.\\
Apartado Postal 4-137
44410 Guadalajara, Jal., M\'exico
\\ \ \\
V.I. Tsifrinovich  \\
 Polytechnic University,\\
 Six Metrotech Center, Brooklyn, NY 11201\\ \ \\
\end{center}
\begin{center}
{\bf Abstract}
\end{center}

The physical implementation of the quantum Control-Not gate for a two-spin
system is investigated numerically. The concept of a generalized 
quantum Control-Not gate, with arbitrary phase shift, is introduced. It is shown that a resonant $\pi$-pulse provides a simple example of a generalized quantum Control-Not gate.
\newpage
Recently, the field of quantum computation has experienced remarkable progress. (See, for example, the review \cite{1}). Important achievements include
a practical 
implementation of quantum computing \cite{2}, a quantum algorithm for prime 
factorization \cite{3}, a quantum error corrections code \cite{4} and an algorithm for pattern recognition \cite{5}. It was shown in \cite{6} that two qubits quantum Control-Not (CN) gates 
in combination with one qubit rotations can provide all the quantum logic 
gate. Because of this, significant attention has been directed toward the implementation of quantum CN gates \cite{7}-\cite{10}. The 
simplest practical implementation of quantum CN logic gates is based on the application  of a $\pi$-pulse to a chain of weakly interacting two-level atoms (ions, spins) which have different 
frequencies of transition \cite{9}. A qubit can be spanned by the ground and long-lived exited
state of an atom. Weak interaction between atoms provides the implementation of the quantum CN gate.

In our previous paper \cite{10}, we explored this idea for the system of two nuclear
spins connected by the Ising interaction. We investigated the evolution of the 
complex amplitudes, $c_{ik}$, of the wave function,
$$
\psi(t)=\sum_{i,k=0}^2c_{ik}(t)|ik>,\eqno(1)
$$
under the action of a resonant $\pi$-pulse. The first number $i$ in $|ik>$ 
indicates the state of the control spin (qubit); the second number $k$  
indicates the state of the target spin (qubit). Our numerical calculations
successfully confirmed the idea \cite{9} concerning the behavior of the moduli of the quantum complex amplitudes. For quantum computations, 
the dynamics of  phases
of complex amplitudes  $c_{ik}$ is also important. Indeed,  quantum interference effects depends on both, moduli and phases. 

In this 
paper, we investigate the dynamics of the quantum CN gate taking into  consideration the behavior of moduli and phases of the complex amplitudes, $c_{ik}$. We consider the system with the Hamiltonian \cite{10},
$$
{\cal H}_0=-{\hbar\over 2}\left( \sum_{n=1}^2\omega_n\sigma_n^z+J\sigma_1^z\sigma_2^z\right).
\eqno(2)
$$
This Hamiltonian describes a four level quantum system with the frequencies: $\omega_n\pm J$. The idea \cite{9} of implementation of a 
quantum CN gate is the following. Assume that one applies to the system (2) a $\pi$-pulse with the frequency, $\omega=\omega_2-J$.  This pulse drives the second (target) spin from the 
ground state $|0>$ to the exited state $|1>$ (or vice versa) if the first 
(control) spin is in exited state $|1>$. We recall that the quantum CN gate
is commonly defined as,
$$
CN=|00><00|+|01><01|+|10><11|+|11><10|.\eqno(3)
$$
It follows from (3) that any superposition state,
$$
\psi=c_{00}(0)|00>+c_{01}(0)|01>+c_{10}(0)|10>+c_{11}(0)|11>,\eqno(4)
$$
after the action of the CN operator (3), is transformed to the state,
$$
\psi^\prime=c_{00}(0)|00>+c_{01}(0)|01>+c_{10}(0)|11>+c_{11}(0)|10>.\eqno(5)
$$
We now introduce the generalized quantum CN gate (GCN gate),
$$
GCN(\Delta\varphi_{ik})=\exp(i\Delta\varphi_{00})|00><00|+
\exp(i\Delta\varphi_{01})|01><01|+\eqno(6)
$$
$$
\exp(i\Delta\varphi_{11})|10><11|+
\exp(i\Delta\varphi_{10})|11><10|.
$$
Similar to a ``pure" CN gate (3), the GCN gate changes the state of the target qubit if the
control qubit is in the state $|1>$. In addition, it changes 
the phases of the complex 
amplitudes $c_{ik}$.  (Up to an insignificant common phase factor, we can put
one of the phases $\Delta\varphi_{ik}$ to be zero.) For quantum 
computations, it is important to know the values of $\Delta\varphi_{ik}$, otherwise one can not describe properly the action of a GCN gate.

Consider the action of a $\pi$-pulse with frequency $\omega$ on a system with the Hamiltonian (2). We add to the Hamiltonian (2) the interaction with a circularly polarized transverse magnetic field,
$$
{\cal H}_{int}=-{\hbar\over 2}\sum_{n=1}^2a_n\left(\exp(-i\omega t)\sigma_n^{-}+
\exp(i\omega t)\sigma_n^{+}\right), \eqno(7)
$$
and substitute the wave function (1) into the Schr\"odinger equation.
  To obtain the equations of
motion with constant coefficients, we use substitution which is equivalent
to a transition to a rotating frame:
$$
c_{00}\to c_{00}\exp[i(\omega t+\varphi)],\eqno(8)
$$
$$
c_{01}\to c_{01}\exp(i\varphi)
$$
$$
c_{10}\to c_{10}\exp(i\varphi),
$$
$$
c_{11}\to c_{11}\exp[i(-\omega t+\varphi)].
$$
In (8), $\varphi=\varphi(t)$ is a common phase (which can be chosen arbitrary, to 
simplify the equations). We shall choose,
$$
\omega=\omega_2-J,\quad \varphi(t)=(\omega_2-J-\omega_1)t/2.\eqno(9)
$$
Then, we derive the following equations (with time-independent coefficients)
for the amplitudes $c_{ik}$, 
$$
-2i\dot c_{00}=-2(\omega_2-\omega_1-2J)c_{00}+a_1c_{10}+a_2c_{01},\eqno(10)
$$
$$
-2i\dot c_{01}=-2(\omega_2-\omega_1)c_{01}+a_1c_{11}+a_2c_{00},
$$
$$
-2i\dot c_{10}=a_1c_{00}+a_2c_{11},
$$
$$
-2i\dot c_{11}=a_1c_{01}+a_2c_{10}.
$$
In matrix representation Eqs (10) have the form,
$$
-2i\dot c_{ik}=B_{ik;jm}c_{jm},\eqno(11)
$$
where $B_{ik;jm}$  is a matrix with time-independent elements. The last two equations in (10) describe the resonant transition between the 
state $|10>$ and $|11>$. The first two equations in (10) describe the nonresonant
dynamics of the lower energy states, $|00>$ and $|01>$. 

For the values of parameters, 
$$
\omega_1=5\omega_2,\quad \omega_2=100,\quad J=5,\quad \omega=\omega_2-J=95,
\quad a_1=0.5,\quad a_2=0.1,\eqno(12)
$$
we obtain the dependence of the complex amplitudes on time during the action of the
$\pi$-pulse. (The characteristic dimensional parameters can be chosen: 
$\omega_2/2\pi=100~MHz$, $\omega_1/2\pi=500~MHz$.) Note that the effective angle of rotation of the second spin under the action 
of the electromagnetic pulse is slightly larger than $a_2\tau$, where
$\tau$ is the duration of a $\pi$-pulse. The reason is a weak indirect excitation
of the resonant transition $|10>\leftrightarrow |11>$ via the nonresonant (first)
spin. The terms $a_1c_{00}$ and $a_1c_{01}$ in the last two equations in (10)
are responsible for this effect.

The free evolution of the two-spin system in the rotating frame 
($a_1=a_2=0$ in equations (10)) is described by the amplitudes,
$$
c_{00}(t)=c_{00}(0)\exp[-i(\omega_2-\omega_1-2J)t],\eqno(13)
$$
$$
c_{01}(t)=c_{01}(0)\exp[-i(\omega_2-\omega_1)t],
$$
$$
c_{10}(t)=c_{10}(0),\quad 
c_{11}(t)=c_{11}(0).
$$
To eliminate the phase factor, corresponding to the free evolution, we will
discuss below the dynamics of the coefficients $ c^\prime_{ik}$,
$$
c^\prime_{00}(t)=c_{00}(t)\exp[i(\omega_2-\omega_1-2J)t],\eqno(14)
$$
$$
c^\prime_{01}(t)=c_{01}(t)\exp[i(\omega_2-\omega_1)t],
$$
$$
c^\prime_{10}(t)=c_{10}(t),\quad
c^\prime_{11}(t)=c_{11}(t).
$$
In Fig. 1a the time dependence of the real part of $c^\prime_{11}$, $ Re c^\prime_{11}(t)$,
and the imaginary part of  $c^\prime_{10}(t)$,  $Im c^\prime_{10}(t)$, are shown for the initial conditions,
$$
c^\prime_{11}(0)=1,\quad c^\prime_{jm}(0)=0, \quad (j,m)\not=(1,1).\eqno(15)
$$
One can see the monotonic decrease of $Re c_{11}^\prime(t)$ and increase of 
$Im c^\prime_{10}(t)$. At the end of the $\pi$-pulse, $c^\prime_{10}(\tau)=
ic^\prime_{11}(0)$.
The values of $Im c^\prime_{11}(t)$ and $Re c^\prime_{10}(t)$ are negligible as well as the values  $|c^\prime_{00}(t)|$ and 
$|c^\prime_{01}(t)|$. This evolution describes the transformation,
$$
|11>\to i|10>.\eqno(16)
$$
In Fig. 1b the analogous dependences are shown for the initial conditions 
$$ 
c^\prime_{10}(0)=1,\quad c^\prime_{jm}(0)=0,\quad (j,m)\not=(1,0).\eqno(17)
$$
At the end of the  $\pi$-pulse, we have: $c^\prime_{11}(\tau)=ic^\prime_{10}(0)$ which
corresponds to the transformation,
$$
|10>\to i|11>.\eqno(18)
$$
For the initial conditions,
$$
c^\prime_{01}(0)=1,\quad c^\prime_{jm}(0)=0,\quad (j,m)\not=(0,1),\eqno(19)
$$
(which corresponds to the population of nonresonant level $|01>$), the amplitudes
$c^\prime_{ik}(t)$ do not change within an accuracy of $10^{-3}$. The same is true for 
the initial conditions, 
$$
c^\prime_{00}(0)=1,\quad c^\prime_{jm}(0)=(0,0),\quad (j,m)\not=(0,0).\eqno(20)
$$
In Fig. 2 the time dependence of the amplitudes is shown for the superpositional
initial state, 
$$
c_{00}^\prime(0)=\sqrt{3/10},\quad
 c_{01}^\prime(0)=1/\sqrt{5},\quad
c_{10}^\prime(0)=1/\sqrt{3},\quad 
c_{11}^\prime(0)=1/\sqrt{6}.
\eqno(21)
$$
One can see that at the end of the $\pi$-pulse, the amplitudes take the 
following values,
$$
c^\prime_{11}(\tau)=ic_{10}^\prime(0),\quad
c_{10}^\prime(\tau)=ic^\prime_{11}(0).\eqno(22)
$$
(The nonresonant amplitudes $c^\prime_{01}(t)$ and $c^\prime_{00}(t)$ do not
change within an accuracy of $10^{-3}$).
One can conclude that a $\pi$-pulse with the frequency $\omega_2-J$ implements a 
GCN gate,
$$
CN(0,0,\pi/2,\pi/2)=|00><00|+|01><01|+i|10><11|+i|11><10|.\eqno(23)
$$
GCN gate (23) can not be used for implementation of a ``pure" CN gate (3). 

We also checked 
an opportunity  for implementation of ``pure" quantum CN gate for the amplitudes $c_{ik}(t)$ in (10), 
which include fast oscillations of the free evolution. This is, probably,
not important for quantum computations, but  may be interesting for experimental investigation of nonresonant quantum states in this system.
We have found that by changing only the amplitude of a $\pi$-pulse, we can not get
a ``pure" CN gate. It was necessary to change slightly one of the parameters
$\omega_1, \omega_2$ or $J$ in (12) to get a ``pure" CN gate.

In Fig. 3 the action of the ``pure" CN gate is demonstrated for the values of
parameters,
$$
\omega_1=500.06,\quad\omega_2=100,\quad J=5,\quad a_2=0.10016,\quad 
a_1={a_2\omega_1\over\omega_2},\eqno(24)
$$
and for ``digital" initial conditions,
$$
c_{ik}(0)=1,\quad c_{jm}(0)=0,\quad (j,m)\not=(i,k),\eqno(25)
$$
where the indices $ik$ take the values $00,01,10$ and $11$ from the top to the
bottom of Fig. 3. Because of the fast oscillations of nonresonant 
amplitudes, we show only the evolution of the amplitudes only near the end of the $\pi$-pulse.
One can see that for any initial ``digital" condition, the only nonzero 
amplitude at the end of a $\pi$-pulse has the value ``i". In Fig. 4 the 
action of the same gate is shown for the superpositional initial conditions (21).
 One can see that at the end of the $\pi$-pulse, 
$$
c_{00}(\tau)=ic_{00}(0),\quad
c_{01}(\tau)=ic_{01}(0),\quad
c_{10}(\tau)=ic_{11}(0),\quad
c_{11}(\tau)=ic_{10}(0).
\eqno(26)
$$
Thus, up to insignificant common phase factor, $\pi/2$, one has
the ``pure" CN gate (3).

We are grateful to D.K. Ferry for fruitful discussions. This research was supported in part by the Linkage Grant 93-1602 from the NATO Special Programme Panel on Nanotechnology. Work at Los Alamos was supported by the Defense Advanced Research Projects Agency.

\vfil\eject

\newpage
\quad\\
{\bf Figure Captions}\\ \ \\
Fig. 1.\quad Time evolution of the amplitudes $c_{ik}^\prime$ under the action of a $\pi$-pulse, for the initial conditions (a) (15) and (b) (17). In (a) curve (1) corresponds to $Re c^\prime_{11}(t)$ and curve (2) corresponds to $Im c^\prime_{10}(t)$. In (b) curve (1) corresponds to $Re c^\prime_{10}(t)$ and curve (2) corresponds to $Im c^\prime_{11}(t)$. The vertical arrows show the beginning and the end of the $\pi$-pulse.\\ \ \\
Fig. 2.\quad  Time evolution of the amplitudes $c_{ik}^\prime$ under the action of a $\pi$-pulse, for the superpositional initial conditions (21).  In (a) curve (1) corresponds to $Re c^\prime_{10}(t)$ and curve (2) corresponds to $Im c^\prime_{10}(t)$. In (b) curve (1) corresponds to $Re c^\prime_{11}(t)$ and curve (2) corresponds to $Im c^\prime_{11}(t)$. The vertical arrows show the beginning and the end of the $\pi$-pulse.\\ \ \\
Fig. 3\quad Imaginary parts of the amplitudes $c_{ik}(t)$ near the end of a $\pi$-pulse for digital initial conditions (25), and the values of parameters (24).\\ \ \\
Fig. 4.\quad Imaginary parts of the amplitudes $c_{ik}(t)$ near the end of a $\pi$-pulse for the superpositional initial conditions (21), and the values of parameters (24).

\end{document}